\documentclass[reprint,groupedaddress,amsmath,amssymb,aps,prl,floatfix]{revtex4-1}

\usepackage{amsmath,amssymb,graphicx,mathrsfs,amsfonts}

\newcommand{\ket}[1]{\ensuremath{\left|  #1 \right\rangle}}

\draft 

\begin{document}

\title{A Cold-Strontium Laser in the Superradiant Crossover Regime} 



\author{Matthew A. Norcia}
\affiliation{JILA, NIST, and University of Colorado, 440 UCB, 
Boulder, CO  80309, USA}
\author{James K. Thompson}
\affiliation{JILA, NIST, and University of Colorado, 440 UCB, 
Boulder, CO  80309, USA}


\date{\today}

\begin{abstract}
Recent proposals suggest that lasers based on narrow dipole-forbidden transitions in cold alkaline earth atoms could achieve linewidths that are orders of magnitude smaller than linewidths of any existing lasers.  
Here, we demonstrate a laser based on the 7.5~kHz linewidth dipole forbidden $^3 $P$_1$ to $^1 $S$_0$ transition in laser-cooled and tightly confined $^{88}$Sr.  We can operate this laser in the bad-cavity regime, where coherence is primarily stored in the atoms, or continuously tune to the more conventional good-cavity regime, where coherence is primarily stored in the light field.  
We show that the cold-atom gain medium can be repumped to achieve quasi steady-state lasing, and demonstrate up to an order of magnitude suppression in the sensitivity of laser frequency to changes in cavity length, the primary limitation for the most frequency stable lasers today.

\end{abstract}


\maketitle

The ongoing quest for lasers with stable and narrow frequency spectra has lead to many advances in both technology and in our understanding of fundamental physics.  Applications include optical clocks \cite{chou2010frequency,Hinkley2013,Bloom14,Katori2015}, searches for gravitational waves \cite{abbott2009ligo}, precision studies of atomic interactions \cite{Zhang19092014}, and tests of the predictions of relativity \cite{chou2010optical, eisele2009laboratory}.

\begin{figure}[!htb]
\includegraphics[width=3.375in]{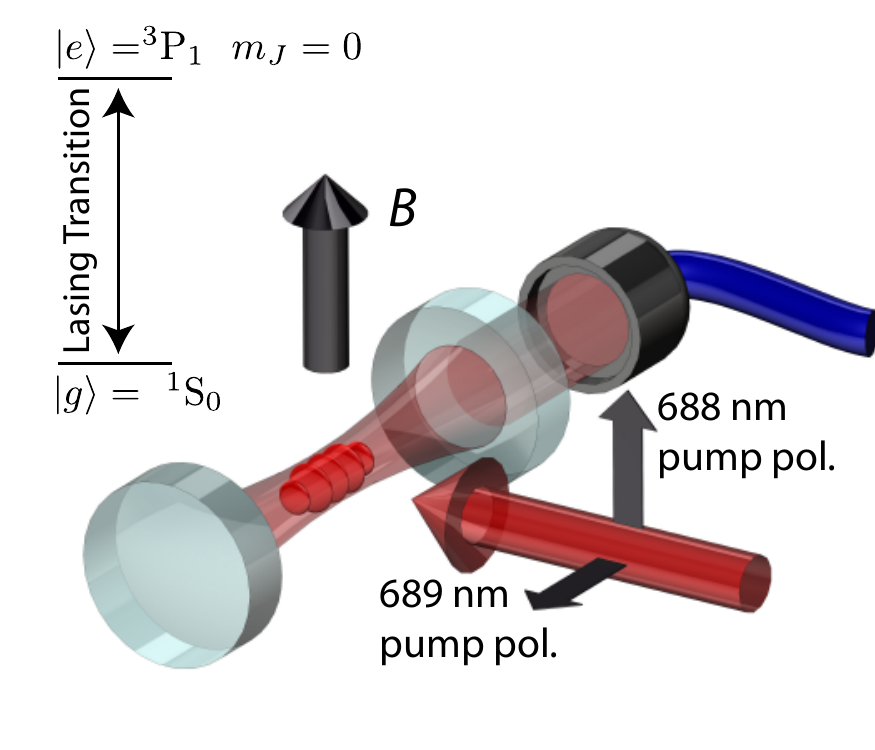}
\caption{ Energy level and experimental diagrams.  The lasing transition is the 7.5~kHz wide $^3$P$_1$ $m_J=0$ to $^1$S$_0$ transition in $^{88}$Sr.  Atoms are confined by a 1D optical lattice in a finesse 24,000 cavity with linewidth $\kappa = 2 \pi~ \times 160$~kHz.  A magnetic field $B$ is oriented perpendicular to the cavity axis, and pump lasers at 688 and 689~nm are applied as shown.  Collectively enhanced emission into the TEM00 mode of the cavity is collected in an optical fiber, and sent to one of various detectors.  }
\label{fig:ExpDiag}
\end{figure}

\begin{figure*}[!htb]
\includegraphics[width=\textwidth]{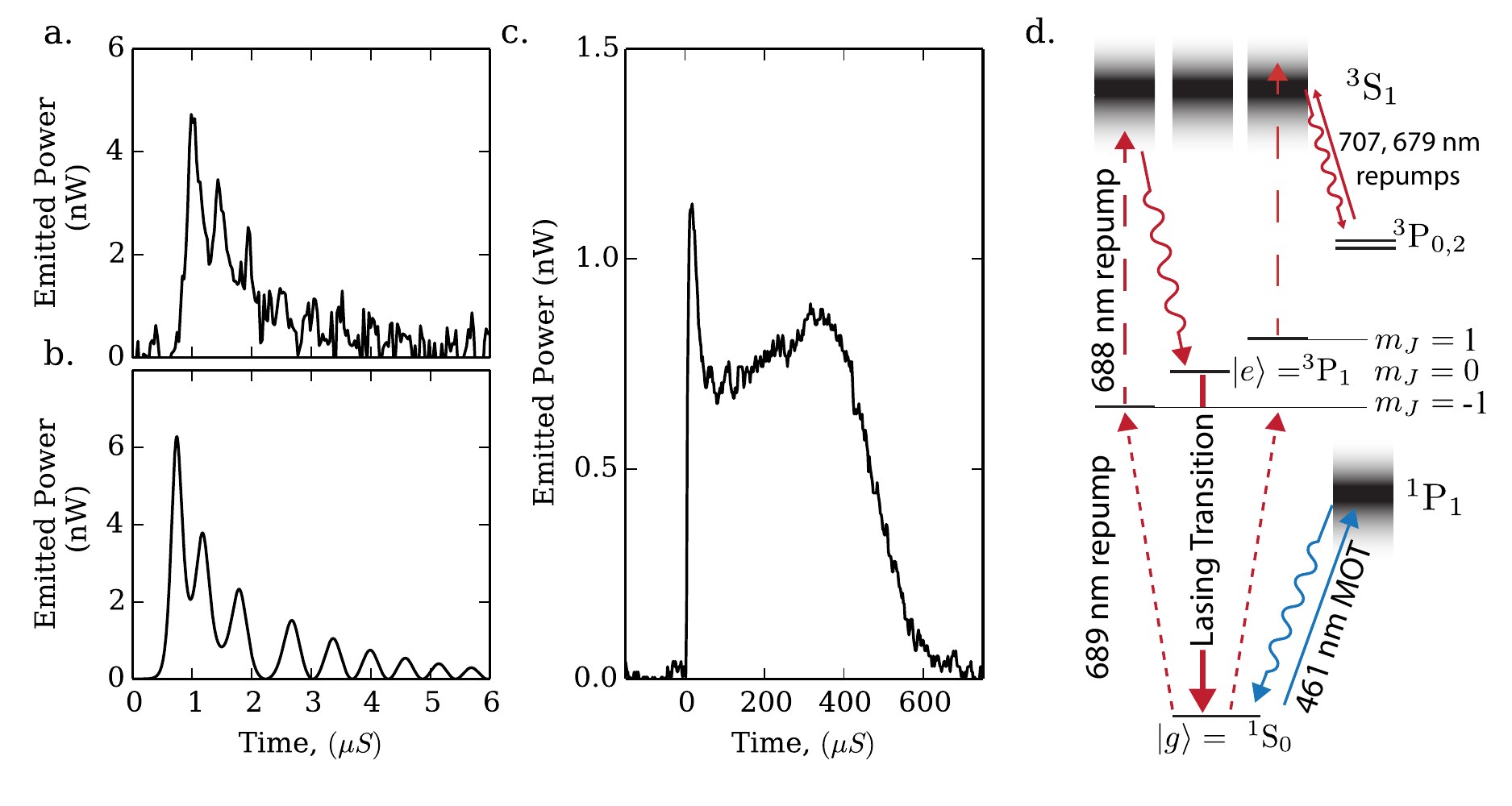}
\caption{(a) Laser power emitted from the cavity during a collectively enhanced pulse, with $N = 36$k atoms.  The atoms are optically pumped into $\ket{e}$ at $t=0.$  (b) Numerical simulation of (a) with no free parameters.  (c) The power emitted from the cavity during continuously repumped lasing with $N = 60$k atoms.  Each atom emits around 35 photons into the cavity before repump-induced heating causes laser operation to cease.  For scale, 1~nW corresponds to 3500 intracavity photons.  (d) Repumping scheme.  Atoms are incoherently repumped through $^3P_1$, $m_J = -1$ to $^3S_1$, where they decay into the $^3P$ manifold.  Atoms that fall into $^3P_0$ or $^3P_2$ are then repumped by additional lasers via $^3S_1$.  For display purposes, $^1$P$_{0,2}$ and $^1$P$_1$ states are shown at arbitrary vertical positions.  
}
\label{fig:ExpDiag}
\end{figure*}

The current state-of-the-art laser technology relies on a broadband gain medium that is frequency stabilized by feedback from a narrow and stable optical cavity.  Deviations in the optical cavity length due to thermal and technical fluctuations are imposed on the laser's frequency by the feedback, and are the primary limitation on the frequency stability of today's most stable lasers \cite{kessler2012sub, numata2004thermal, notcutt2006contribution}.  An alternative approach is to operate in a bad-cavity, or superradiant regime, where a narrow-band gain medium is confined within a relatively lossy optical cavity, analogous to maser operation in the microwave domain.  In a laser operating in the superradiant regime, the frequency of the emitted light is primarily set by the transition frequency of the gain medium, rather than the optical cavity.  
Recently, it has been proposed that one could operate such a laser based on highly forbidden transitions in cooled and trapped alkaline  earth atoms, such as the mHz wide $^3$P$_0$ to $^1$S$_0$ transition in $^{87}$Sr, to achieve a linewidth of 1~mHz or narrower \cite{MYC09, MEH10, CHE09}.

Here, we take a step towards realization of such a frequency standard by demonstrating and characterizing lasing on the $7.5$~kHz wide dipole-forbidden $^3$P$_1$ to $^1$S$_0$ transition at $689$~nm in an ensemble of $^{88}$Sr tightly trapped in a 1D optical lattice.  This laser operates at the crossover  between good- and bad-cavity regimes, and can be tuned between the two by varying homogeneous broadening of the lasing transition.  We demonstrate that this laser can be repumped, resulting in quasi-steady-state-operation, and show that the frequency of the emitted light is set primarily by the atomic transition frequency when operated in the bad-cavity regime, and by the optical cavity resonance frequency in the good-cavity regime.

The bad-cavity regime of laser physics has been explored in the optical domain in gas lasers \cite{PhysRevLett.72.3815}, with homogeneous and inhomogeneous transition linewidths of hundreds of MHz, and in a 4-level system in Cesium with an Doppler-broadened gain bandwidth of 9~MHz \cite{xu2014lasing}.  In the microwave domain, masers operate deep in the bad-cavity regime \cite{GKR60}. Rare-earth doped solid state lasers utilize transitions with long-lived excited states, but with inhomogeneous broadening of hundreds of GHz \cite{siegman1986lasers}.
Raman dressing has been used to create long-lived virtual states with low inhomogeneous broadening, which have been used to  demonstrate key properties of optical lasing very deep into the bad-cavity regime \cite{bohnet2012steady, BCWHybrid, BCWDyn}, and in the deep good-cavity regime \cite{PhysRevLett.107.063904}.  The results presented here explore lasing on a true optical transition in a regime where both the homogeneous and inhomogeous linewidths of the gain medium can be made small compared to the cavity linewidth. This is a key step towards a useful frequency reference based on even more narrow transitions \cite{MYC09, MEH10, CHE09, Maier14, kazakov2015active}.


Our system, also described in \cite{norcia2015strong}, consists of up to $N \sim 100 $k $^{88}$Sr atoms cooled to 9 $\mu$K and confined by an optical lattice within a high finesse optical cavity (Fig.~1).  
Atoms are prepared via a two-stage cooling process, with initial capture and cooling using the dipole-allowed $^1$S$_0$ to $^1$P$_1$ transition at 461~nm, and final cooling and lattice loading using the narrow $^1$S$_0$ to  $^3$P$_1$ transition.  
The 813~nm lattice is supported by a TEM00 mode of the cavity, which provides intensity buildup for the lattice, and precise spatial registration of the atoms with respect to the cavity mode.  
The lattice laser and cavity resonance frequencies are stabilized relative to a separate spectroscopy signal from the $^1$S$_0$ to $^3$P$_1$, $m_J = 0$ transition, allowing for precise tuning of the cavity resonance frequency.

At the lasing frequency of 689~nm, the cavity has a finesse of 24,000 and a linewidth $\kappa = 2 \pi \times 160$~kHz.  
The lasing occurs on the $\ket{e}\equiv\ket{^3\mathrm{P}_1, m_J = 0}$ to $\ket{g}\equiv\ket{^1\mathrm{S}_0}$ transition at 689~nm, which has a natural decay linewidth $\gamma = 2\pi \times 7.5$ kHz.
The atomic coupling to the cavity mode is inhomogeneous in strength because the lasing and trap wavelengths differ significantly.
Atoms trapped at an antinode of the lasing mode exchange excitations with the TEM00 mode at a frequency $2g_0 = 2 \pi \ \times$ 21.2 kHz.  Accounting for inhomogeneous coupling, this frequency is collectively enhanced to up to $\Omega = g_0 \sqrt{2 N}\sim 2\pi\times1~ \mathrm{to}~5$~MHz.  
Our system operates in the single particle weak coupling regime, $C = \frac{4 g_0^2}{\kappa \gamma} = .41(4) < 1$, but in the collective strong coupling regime, $NC \gg 1$.

When many atoms are placed in state $\ket{e}$, the collectively enhanced coupling to the cavity causes the atoms to quickly decay to $\ket{g}$ by emitting a pulse of light into the cavity.  We detect this light on an avalanche photodiode, with a single trace shown in Fig.~2a.  The collectively enhanced emission rate exceeds the cavity linewidth, which in turn is much greater than the atomic decay rate, i.e.~ $\Omega\gg\kappa\gg\gamma$.  In this regime, the light-pulse is partially re-absorbed by the atoms, and re-emitted into the cavity several times before escaping the system via transmission through a cavity mirror, resulting in oscillations in output power observed in both the data and simulation of Fig 2a and b.  The timescale of these oscillations is of order $2\pi/\Omega$, but because of inhomogeneous coupling to the cavity, the atoms do not remain in a fully symmetric state.  This modifies the period of oscillations, and results in incomplete reabsorption of the initially emitted pulse.  

By measuring a vacuum Rabi splitting following the pulse, as described in \cite{norcia2015strong}, we infer an atom number of $N = 36,000$ for the trial shown in Fig.~2a, up to fluctuations between trials of around 20 percent.  We simulate the pulse dynamics for this atom number by integrating a set of optical Bloch equations \cite{bohnet2014linear} that also account for the inhomogeneous coupling to the cavity mode.  The results are shown in Fig.~2b, with good qualitative agreement to observed pulses.  


We next apply continuous repumping from $\ket{g}$ back to $\ket{e}$ to operate the laser in a steady-state manner.  The repumping process both maintains population inversion, and causes decay of the transverse coherence of the atomic ensemble, which reduces the collectively enhanced emission rate.  When a steady state operating condition exists, the total rate of emission from $^3$P$_1$ matches the single-particle rate $w$ at which atoms are optically pumped out of $\ket{g}$.  The repumping process also homogeneously broadens the ground state to a width $w$.  When operating with $w \ll \kappa$, we access the superradiant regime, where the cavity decay rate exceeds the collectively enhanced emission rate, other damping rates, and the bandwidth of the gain medium.

In Fig.~2c, we show the laser output power for a representative run of the experiment, with repumping applied to an ensemble of $N \simeq 60$k atoms initially in $\ket{e}$.  We can sustain lasing for up to 1.5~ms, with higher power operation possible for durations of around 500~$\mu$s, as shown.  
In such a trial, each atom emits up to 35 photons before being lost due to photon recoil heating.

The repumping mechanism is shown in Fig.~2d, with beam directions and polarizations shown in Fig.~1a.  A magnetic field is applied perpendicular to the cavity axis, splitting out the $m_J = \pm 1$ states by $\pm 7$ MHz.  
The 689~nm repump light is applied near resonance with the $^1 S _0$ to $^3 P _1$, $m_J = -1$ transition.
Additional $\pi$-polarized 688~nm repump light resonant with the dipole-allowed $^3 P_1$ to $^3 S_1 $ transition is applied from the same direction. 
Because of dipole selection rules, this 688~nm pump only couples the $^3 P _1$ $m_J = \pm 1$ states to $^3 S_1 $, but the $m_J = 0$ state is unaffected.  
Single particle spontaneous emission takes atoms from the $^3S_1$ states to the $^3P$ states, and additional repump lasers at 707~nm and 679~nm are applied to depopulate the metastable $^3 P_2$ and $^3 P _0$ states.  
We emphasize that because atoms reach $^3$P$_1$ by single-particle spontaneous decay, we do not expect any coherence between the pump lasers and the emitted light.  

We believe the lasing terminates when heating from free-space scattering caused by the repump process reduces the product $N C$ both by causing atom loss, and by reducing the coupling of remaining atoms to the cavity.  For each lasing photon emitted into the cavity, we estimate roughly 13 photon recoils are imparted to each atom during the repump process.  For 35 lasing photons, this would contribute heating comparable to the lattice depth.   

In contrast to gas and solid state lasers that operate on similarly long-lived transitions, our laser also operates in a regime of low inhomogeneous broadening. The frequency scale to which inhomogeneous broadening should be compared is the collectively enhanced emission rate, which in the steady state case is set by $w$ \cite{2015arXiv150306464W, PhysRevLett.113.154101}.  
When inhomogeneous broadening is comparable to or less than the collectively enhanced emission rate, the majority of atoms in the ensemble can phase-synchronize and contribute to collectively enhanced emission.  

In our system, Doppler broadening is suppressed because the atoms are confined to much less than the optical wavelength along the cavity axis.  However, the lattice contributes a polarization-dependent shift of the lasing transition \cite{PhysRevLett.91.053001}.  The degree to which this shift is inhomogeneous depends on the temperature of the atoms, which increases during laser operation.  At the beginning of laser operation, we estimate inhomogeneous broadening to be less than 30~kHz FWHM for our typical lattice polarization.  In principle, this broadening can be tuned to zero.  For our range of operating parameters, the majority of atoms have transition frequencies within $w$ of the average transition frequency, allowing them to contribute to lasing.  

\begin{figure}[!htb]
\includegraphics[width=3.55in]{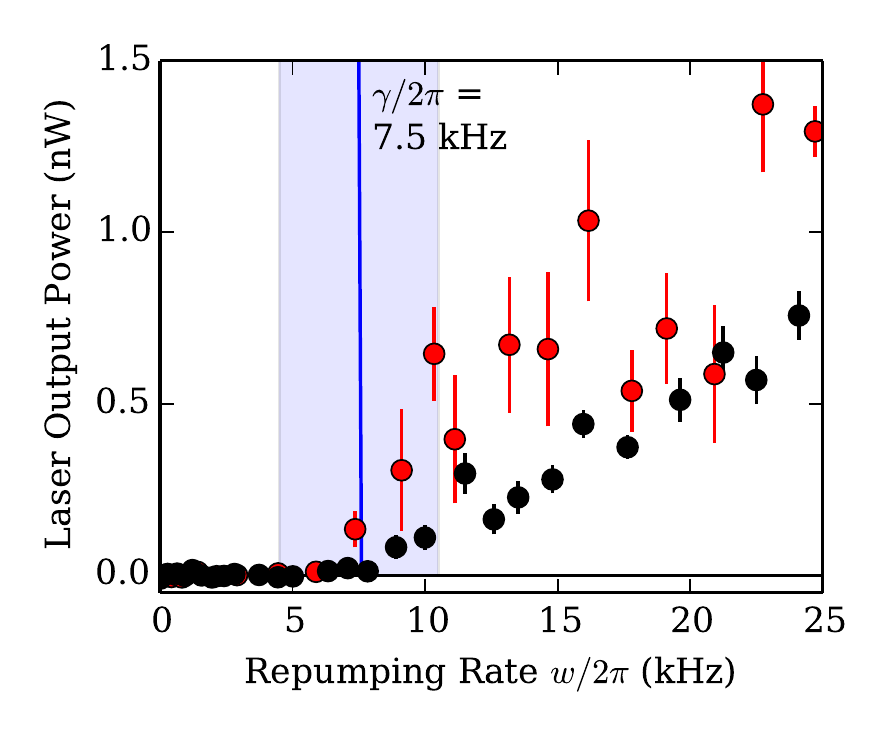}
\caption{Threshold behavior of laser.  Laser output power in quasi-steady state is plotted versus repumping rate $w$ for no added broadening $\gamma^\prime_\perp = 0$ (red points) and added broadening $2\gamma^\prime_\perp \simeq 2 \pi \times~3~\mathrm{MHz} \gg \gamma, \kappa$ (black points).  Atomic decay rate $\gamma$ is displayed as vertical blue line, with shaded blue region representing uncertainty in calibration of $w$. For both conditions, the measured threshold repuming rate $w_t$ is consistent with $w_t = \gamma$.  }
\label{fig:ExpDiag}
\end{figure}

\begin{figure*}[!htb]
\includegraphics[width=6.5in]{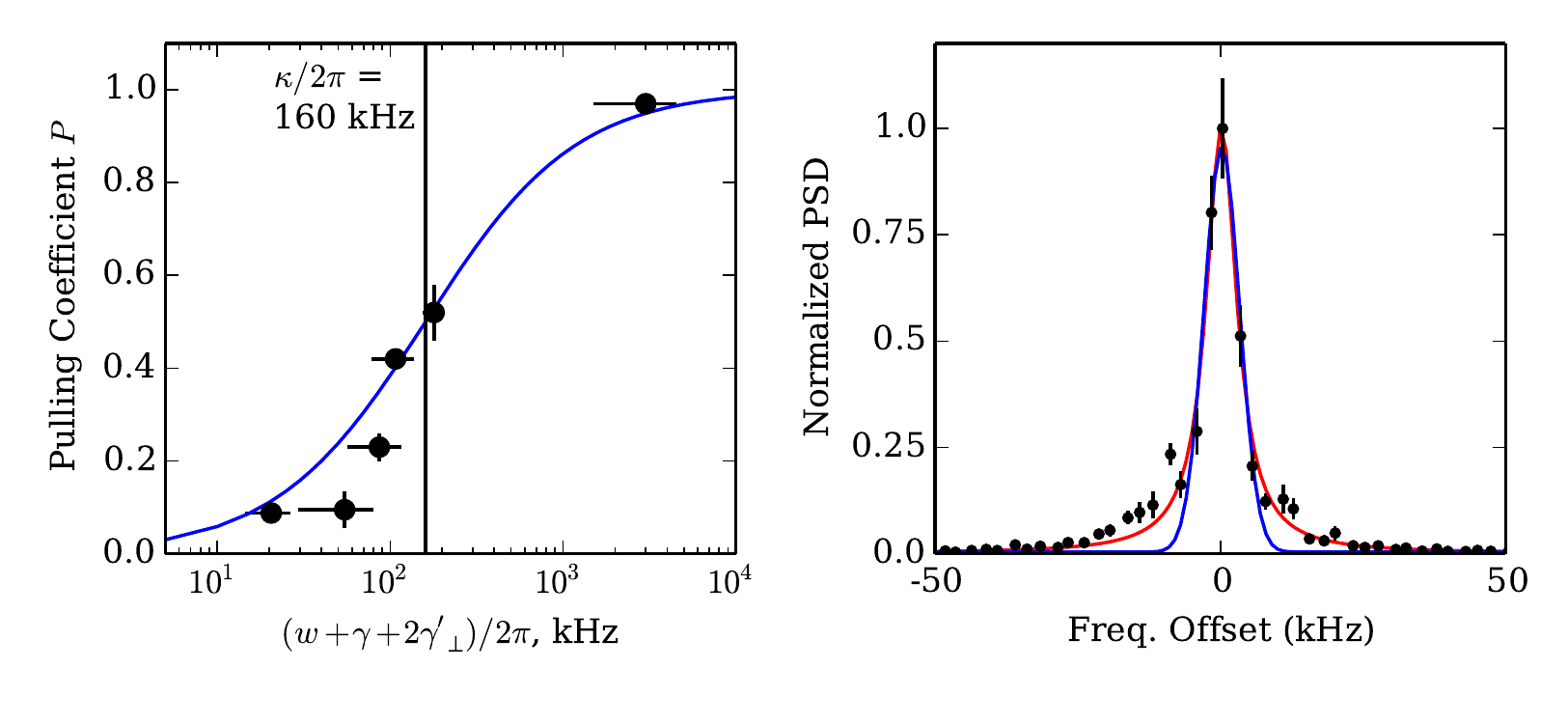}
\caption{ (a) Pulling coefficient $P$ versus the total homogeneous broadening of the lasing transition, along with prediction (blue).  The vertical line at 160~kHz represents the cavity decay rate $\kappa$.  (b) Averaged heterodyne power spectral density (PSD) between output light and 689~nm pump laser, with recentering of each individual trial before averaging.  Lorentzian (gaussian) fits, shown as red (blue) lines indicate FWHM linewidths of 6.0(3) (4.7(3)) kHz.  }
\label{fig:ExpDiag}
\end{figure*}

The laser output power exhibits a characteristic threshold behavior versus $w$, as shown in Fig.~3a.  
In the regime that the total atomic transverse decoherence rate $\gamma_\perp = \gamma/2 + w/2 + \gamma^\prime_\perp$ satisfies $\gamma_\perp \ll N C \gamma$, the threshold repumping rate for lasing is simply $w_t=\gamma$.  This is expected to be true even in the presence of homogeneous or inhomogeneous broadening that exceed the natural decay rate of the lasing transition \cite{MYC09}.  In the above, $\gamma^\prime_\perp$  accounts for other mechanisms that lead to a decay of atomic coherence.  Intuitively, this threshold requirement states that in order to establish population inversion, atoms must be removed from $\ket{g}$ at a rate faster than they can undergo single particle decay from $\ket{e}$ to $\ket{g}$.

We measure output power from the laser in a window between 20 and 60~$\mu$s after the beginning of laser operation.  We tune $w$ by changing the intensity of the 689~nm repump, and measure $w_t = 2 \pi ~\times~  8(3)$~kHz \cite{ThreshFitRefNote}, with the error dominated by the calibration of $w$.  This is consistent with the predicted threshold $w_t$=$\gamma = 2 \pi\times~7.5$~kHz.



By inducing Rayleigh scattering on the dipole-allowed $\ket{g}$ to $^1$P$_1$ transition at 461~nm using our MOT beams, we add additional homogeneous broadening $2\gamma^\prime_\perp \simeq 2 \pi~\times ~3$~MHz $\gg\kappa, \gamma$.  
Despite this high scattering rate, we measure the same threshold value of $w_t =2 \pi~ \times ~8(3)$~kHz, again consistent with the prediction.  
By rotating the lattice polarization, we can introduce inhomogeneous broadening that can be varied from less than 30~kHz to as much as 140~kHz FWHM.  Between these two conditions, we measure $w_t$ to differ by less than 3~kHz.  
This confirms that the dominant factor in determining threshold is indeed the bare atomic decay rate, not the effective transition linewidth.



An important technological appeal of bad-cavity lasers is their reduced sensitivity to fluctuations in cavity resonance frequency.  We define the pulling coefficient $P = \Delta f_l / \Delta f_c$, where $\Delta f_l$ is the change in the lasing frequency created by a change $\Delta f_c$ is the cavity resonance frequency.  We expect the pulling coefficient to be given by $P = 2\gamma_{\perp}/(2\gamma_{\perp}+\kappa)$ \cite{PhysRevLett.72.3815, bohnet2012steady, BCWDyn}.  A pulling coefficient $P \ll 1/2$ is a key signature that the laser is operating in the bad-cavity, or superradiant regime with coherence primarily stored in the atoms, not in the light field.  

To measure the pulling coefficient, we overlap the laser light emitted from the cavity with a heterodyne beam from a frequency stabilized 689~nm laser.  We change the cavity frequency between trials, and compute power spectra from the heterodyne data.  From gaussian fits to these power spectra, we extract the peak frequency of the emitted light.  We obtain the pulling coefficient from a linear fit to $f_l$ versus $f_c$, with $f_c$ scanned by 1~ MHz.  We repeat this with different values of $w$.  
When we turn down the repump beams to just above threshold, $w = 2 \pi \times 14$ kHz, we measure a pulling coefficient of $P=0.09(2)$.  This indicates that we have reached the bad-cavity regime, where the spectral properties of the emitted light are dominated by the atomic gain medium.  When we increase the repump rate such that the effective atomic linewidth becomes comparable to the cavity decay rate, we reach a crossover regime where coherence is shared between atoms and light field, and measure a pulling coefficient consistent with $P=1/2$.  

The right-most point in Fig.~4a was obtained by turning on the 461~nm MOT beams to further increase $\gamma_\perp$ by Rayleigh scattering from $\ket{g}$ at a rate $2\gamma^\prime_{\perp} \simeq 2 \pi \times 3 \mathrm{ ~MHz}$.  With this scattering rate, we measure a pulling coefficient of $P= 0.97(3).$
In this highly broadened regime, scattering from the 461~nm MOT beams removes coherence between $^1S_0$ and $^3P_1$ without directly affecting the population inversion.  By removing the phase coherence from the atoms, we access the good-cavity regime, where the spectral properties of the emitted light are set by the optical cavity.

We bound the linewidth of the emitted light by computing an average power spectrum from heterodyne data taken over many trials.  From trial to trial, the center frequency of the power spectrum fluctuates due to low frequency noise on the heterodyne laser and optical cavity.  We align each power spectrum by shifting its frequency axes by the center frequency obtained from a gaussian fit, which reduces sensitivity to low-frequency noise.  We then fit a lorentzian and gaussian to points from all recentered spectra simultaneously.  From 150~$\mu$s long subsets of the time data, we measure a lorentzian (gaussian) FWHM linewidth of 6.0(3) (4.7(3))~kHz.  We expect that the fundamental quantum linewidth for these operating conditions would be around 1~Hz, \cite{PhysRevLett.72.3815} far below our ability to currently resolve.   Our measured linewidth is fourier limited for shorter subsets, and limited by acoustic noise on the heterodyne laser for longer subsets.  However, our measured linewidth is slightly narrower than the natural linewidth of the lasing transition (7.5~kHz), and far narrower than the linewidth imposed by repumping ($\sim 100$ kHz), exhibiting the frequency narrowing characteristic of synchronization in a laser.


We have demonstrated lasing on a dipole-forbidden transition in cold strontium atoms in both a pulsed mode, and in quasi-steady state.  We can operate this laser in the bad-cavity regime, where the frequency of the emitted light is set primarily by the atomic transition, with sensitivity to cavity frequency shifts reduced by a factor of ten.  By applying these same techniques to the even narrower $^1$S$_0$ to $^3$P$_0$ transition in $^{87}$Sr, one may be able to realize a frequency standard with linewidth well below the current state of the art, and with many orders of magnitude lower sensitivity to environmental vibration noise, which could allow it to be used outside of a laboratory environment.

We acknowledge contributions to the experimental apparatus by Matthew Winchester, and thank Kevin Cox, Jan Thomsen, Murray Holland and David Tieri for useful discussions.  
All authors acknowledge financial support from DARPA QuASAR, ARO, NSF PFC, and NIST. This work is supported by the National Science Foundation under Grant Number 1125844.

\bibliographystyle{apsrev4-1}
\bibliography{ThompsonLab.bib}

\begin{thebibliography}{30}
\bibitem [{Thr()}]{ThreshFitRefNote}%
  \BibitemOpen
  \href@noop {} {\ }\bibinfo {note} {Threshold is determined by fitting to
  output power versus $w$ a function whose value is zero below a threshold
  value $w_t$, and $a(w-w_t)^b$ above.}\BibitemShut {Stop}%
\end{thebibliography}

\end{document}